\documentclass[12pt,draftclsnofoot,onecolumn]{IEEEtran}
\usepackage{amssymb}
\usepackage{amsfonts}
\usepackage{amsfonts,subfigure,multicol,color,verbatim,
graphicx,cite,epsfig,amssymb,amsmath,cases,bm,algorithm,
algorithmic,xcolor,multirow,array}
\usepackage{epstopdf}
\begin{document}
\markboth{China Communications, Invited Paper, vol. XX, no. Y, Month
2016} {Peng: MC \ldots}

\title{Diffusion Based Molecular Communication: Principle, Key Technologies, and Challenges}
\author{Jiaxing~Wang, Bonan Yin, and Mugen~Peng
\thanks{Jiaxing~Wang (e-mail: {\tt jx19882008@163.com}), Bonan Yin (e-mail: {\tt 2402637853@qq.com}), and Mugen~Peng (e-mail: {\tt pmg@bupt.edu.cn}) are with the Key Laboratory of Universal Wireless Communications for Ministry of Education, Beijing University of Posts and Telecommunications, Beijing, China.}
}

\maketitle

\begin{abstract}
Molecular communication (MC) is a kind of communication technology based on biochemical molecules for internet of bio-nano things, in which the biochemical molecule is used as the information carrier for the interconnection of nano-devices. In this paper, the basic principle of diffusion based MC and the corresponding key technologies are comprehensively surveyed. In particular, the state-of-the-art achievements relative to the diffusion based MC are discussed and compared, including the system model, the system performance analysis with key influencing factors, the information coding and modulation techniques. Meanwhile, the multi-hop nano-network based on the diffusion MC is presented as well. Additionally, given the extensiveness of the research area, open issues and challenges are presented to spur future investigations, in which the involvement of channel model, information theory, self-organizing nano-network, and biochemical applications are put forward.
\end{abstract}

\newpage

\section{Introduction}

Molecular communication (MC) is a bio-inspired technology to enable the traditional communication in nano-micro scale environment,
which has stimulated a great deal of interest in both academic and industrial community.
With the development of nano-technology and bioengineering technology, the MC based body area nano-network (BANN) began to be investigated to attempt to realize practical nano-medical applications {\cite{b1}}, which is shown in Fig. 1. In order to handle complex medical tasks in BANNs, the nano-machines with simple calculation, storage, sensing, and communication capabilities should interconnect with each others through the MC technology. Therefore, the incompatibility and possible damage in the body area can be significantly reduced to a minimum level through MC and the synthetic biology {\cite{b2}}. It is generally believed that MC is one of the most feasible communication technologies to realize BANNs for the internet of bio-nano things{\cite{b3}}.
\begin{figure}[H]
 \centering
 \includegraphics[width=0.85\textwidth]{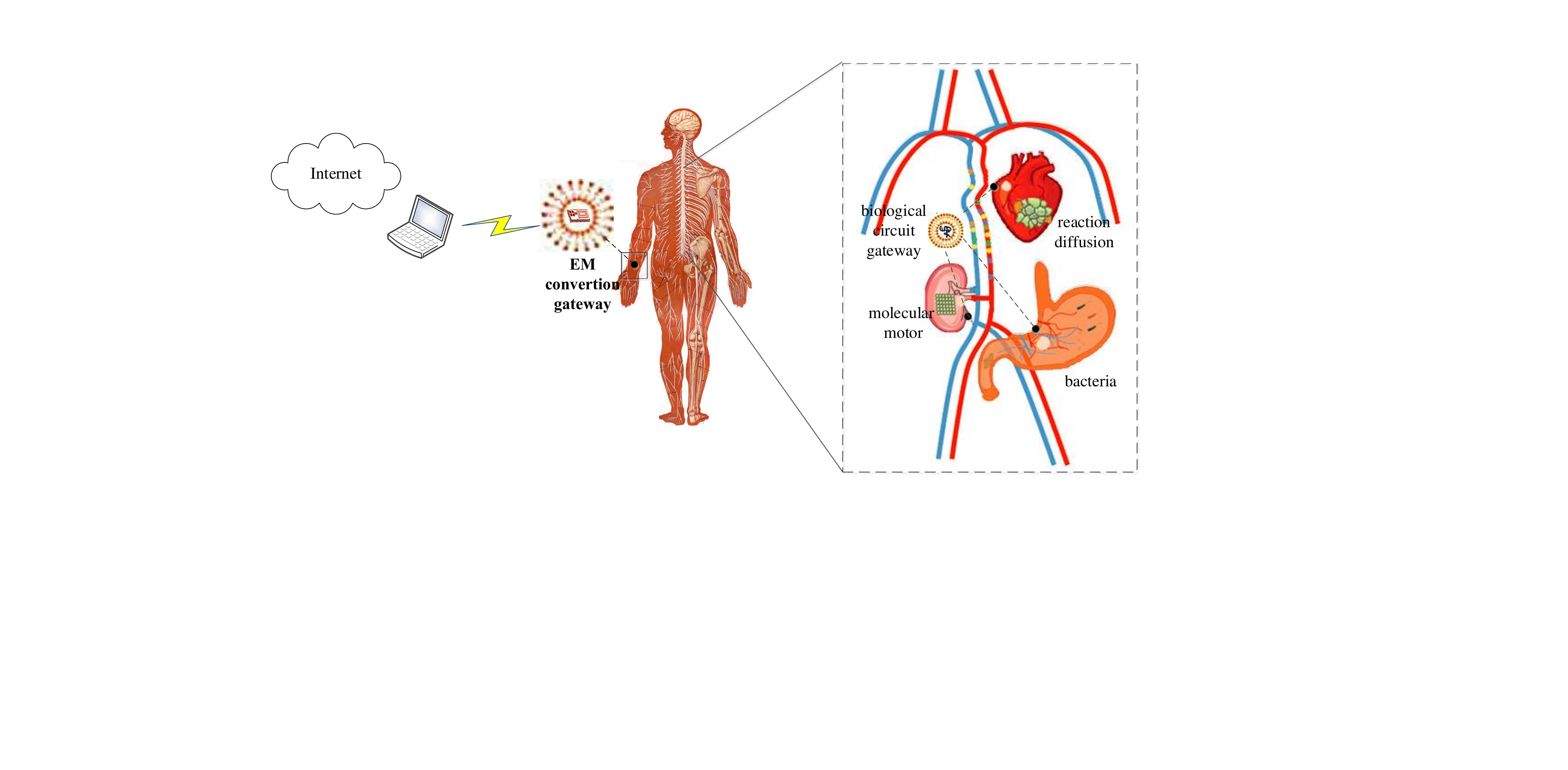}
 \vspace*{-10pt} \caption{System model for the Body-area nano-network (BANN).} \vspace*{-10pt} \label{1}
\end{figure}

MC is a promising but cross research field, which involves a mass of disciplines and technologies, such as the biology (including molecular biology, systems biology and synthetic biology), chemistry, physics, computer science, communication engineering and electronic engineering. Current researches mainly focus on the MC system design and realization, the MC theoretical analysis and nano-networking, namely related to the communication engineering, information theory and computer networking three academic sectors.

MC can be regarded as a kind of micro- or nano-scale communication technology using biochemical molecules as the carrier of information {\cite{b4}}, in which the principles and mechanisms of MC are presented from the microscopic biological communication process in nature. The detailed comparison between MC and the traditional electromagnetic or acoustic wave communication technologies are shown in Table I {\cite{b5}}{\cite{b6}}.
\begin{table*}[tbp]
\centering
\caption{Characteristics comparison between MC and the traditional communication technologies}
\begin{tabular}{|p{5cm}|p{5cm}|p{5cm}|}
\hline
Communication Features &Molecular Communication &Traditional Communication\\ \hline
Information Carrier &Molecule &Electromagnetic or acoustic wave\\\hline
Signal Type &Chemical signals &Electromagnetic or acoustic signals\\\hline
Transmission Medium &Aqueous &Space or cable\\\hline
Noise Source &Particle or chemical reaction in the medium &Electromagnetic field or other signal\\\hline
Transmission Device &Bio-nano-machine &Electronic device\\\hline
Transmission Speed &nm/s or um/s &Speed of light or sound\\\hline
Transmission Distance &nm $\sim$ m &m $\sim$ km\\\hline
Information Content &Chemical state or digital information & Text, video or audio\\\hline
Energy Consumption &Low &High\\ \hline
\end{tabular}
\end{table*}

Compared with the traditional communication technologies, MC has severe propagation delay and the corresponding delay rapidly increases with the arising communication distance, which is a tough challenge for the information retransmission and system synchronization. In addition, due to the molecular diffusion and molecular decay, the communication range of MC is strictly limited by the channel attenuation. Meanwhile, the stochastic nature of Brownian motion and the unpredictable chemical reaction produce a large channel distortion, which limits the minimum time interval between two messages transmission and creates a severe inter-symbol interference (ISI). Even though these challenges, there are still significant benefits in MC against the traditional communication technologies.

\begin{enumerate}
  \item Nano-scale communication: MC is feasible to accomplish a mass of emerging medicine monitor and therapy applications in body area environment. MC makes the operation precise and is expected to realize the targeted and noninvasive therapy.
  \item Biological compatibility: The biologically inspired MC mechanism enables the nano-machines to interact directly with various native components (e.g., cells, tissues, organs, and etc.) in the biological system, which can avoid nano-machines or external stimulations are introduced {\cite{b7}}.
  \item High energy efficient: MC can obtain sufficient energy {\cite{b8}} from the chemical reactions in the body area environment to support the information transmission. For example, myosin (a tool carried information molecules) can use the chemical energy for mechanical operations with the efficiency of nearly 100\%.
\end{enumerate}

The characteristics of MC can be identified by the theoretical research. The substantial researches have been focused on the performance optimization especially for MC, which are inherited from the traditional communication techniques. These researcher have promoted MC mature and available to be implemented in practice. Meanwhile, the related MC standard work in general terms and protocols are being carried out by IEEE COM/Nano-scale and molecular communication working group, which will effectively promote the MC's development and application.

Although initial efforts have been made on the survey of MCs, limitations exist in the previous works that the content is only related to the recent advancements in the field of MC engineering in {\cite{survey9}}, where the biological, chemical, and physical processes used by MC are discussed and presented in a tutorial fashion. Considering MC from the the field of communication and networking engineering, a more comprehensive survey framework for incorporating the basics and the latest achievements on the system model and the corresponding key techniques for the diffusion based MC seems
timely and significant.

The rest of the paper is organized as follows. Section II describes the principle of MC. The performance analysis of diffusion based MC is presented in Section III. Section IV provides the information coding and modulation techniques. Section V describes the multi-hop nano-networks based on the diffusion MC. The open issues and potential challenges are shown in Section VI. Finally, the paper is concluded in Section VII.

\section{Principle of Molecular Communication}

For MC, the information sender nano-machines (transmitter) generates and encodes information into molecules based on the molecular physical or chemical properties, where the channel environment can be fluid (liquid or gas) transmission medium. The receiver receives and decodes information through triggering the chemical reaction or the changes of the chemical state information. With respect to the electromagnetic communication, the natural MC phenomena is commonly found among the nano-bio-entities, such as the communication among cells and bacteria. These natural phenomena provide existing transmission mechanisms, which lay a great good foundation for the investigation of MC.

\subsection{Transmission Model}

As shown in Fig. 2, the basic process of MC includes five main steps, namely, coding, transmission, propagation, reception and decoding of information. Coding is the process that the transmitter generates information molecules with special physical or chemical properties. Transmission is the process that the information molecules are released into the transmission medium. Propagation is the process that information molecules are transmitted from the transmitter to the receiver through a fluid medium. Reception is the process that the information molecules are detected by the receiver. Decoding is the process that the receiver reconstructs information extracted from the detected molecules.

\begin{figure}[H]
 \centering
 \includegraphics[width=0.85\textwidth]{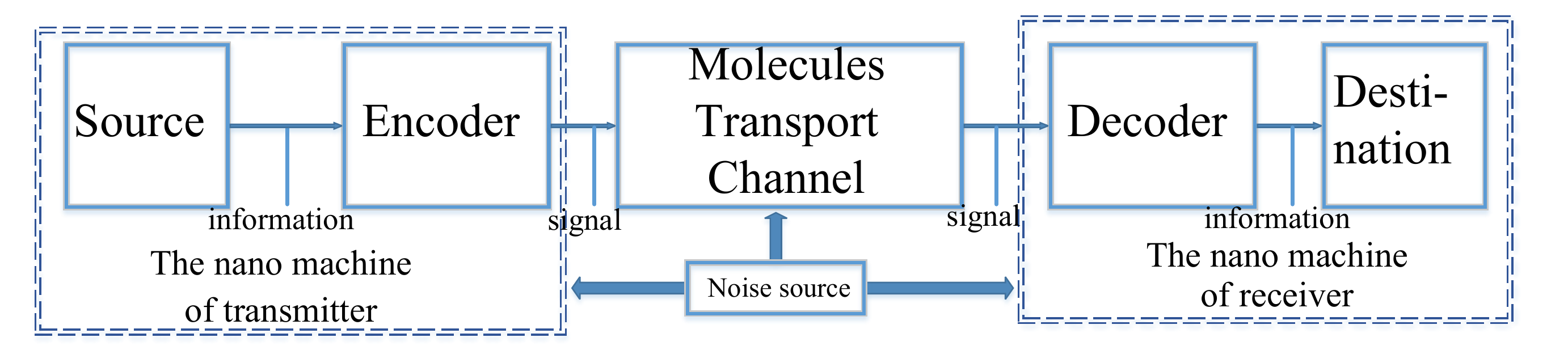}
 \vspace*{-10pt} \caption{The transmission model for MC} \vspace*{-10pt} \label{2}
\end{figure}

\subsection{Transmit Communication Mechanism }

Various transport mechanisms for MC are found within or between cells, which can be categorized based on how information molecules are propagated, namely, whether information molecules directionally propagate by consuming chemical energy or they simply diffuse in the environment. The former type is called active transport mechanism, while the latter is called passive transport mechanism.

\subsubsection{Active Transport Mechanism}

The active transport mechanism consumes chemical energy and generates sufficient power to provide directional transmission towards the specific locations, which can guarantee a high degree of reliability even if the number of information molecules executing communication is small. However, the communication infrastructure is necessary for active transport mechanism to maintain the transmission energy and guiding. The active transport mechanism generally uses molecular motors or bacteria.

$a)$ Molecular motors transport mechanism. The molecular motors consume chemical energy to transport information molecules or the vesicles along the microtubules {\cite{b9}}, where the molecular motor is a protein or a protein complex that moves along a microtubule unidirectionally according to the polarity of the microtubule, e.g., dynein, kinesin. The molecular motor based MC transmission schematic diagram is shown in Fig. 3.

\begin{figure}[H]
 \centering
 \includegraphics[width=0.6\textwidth]{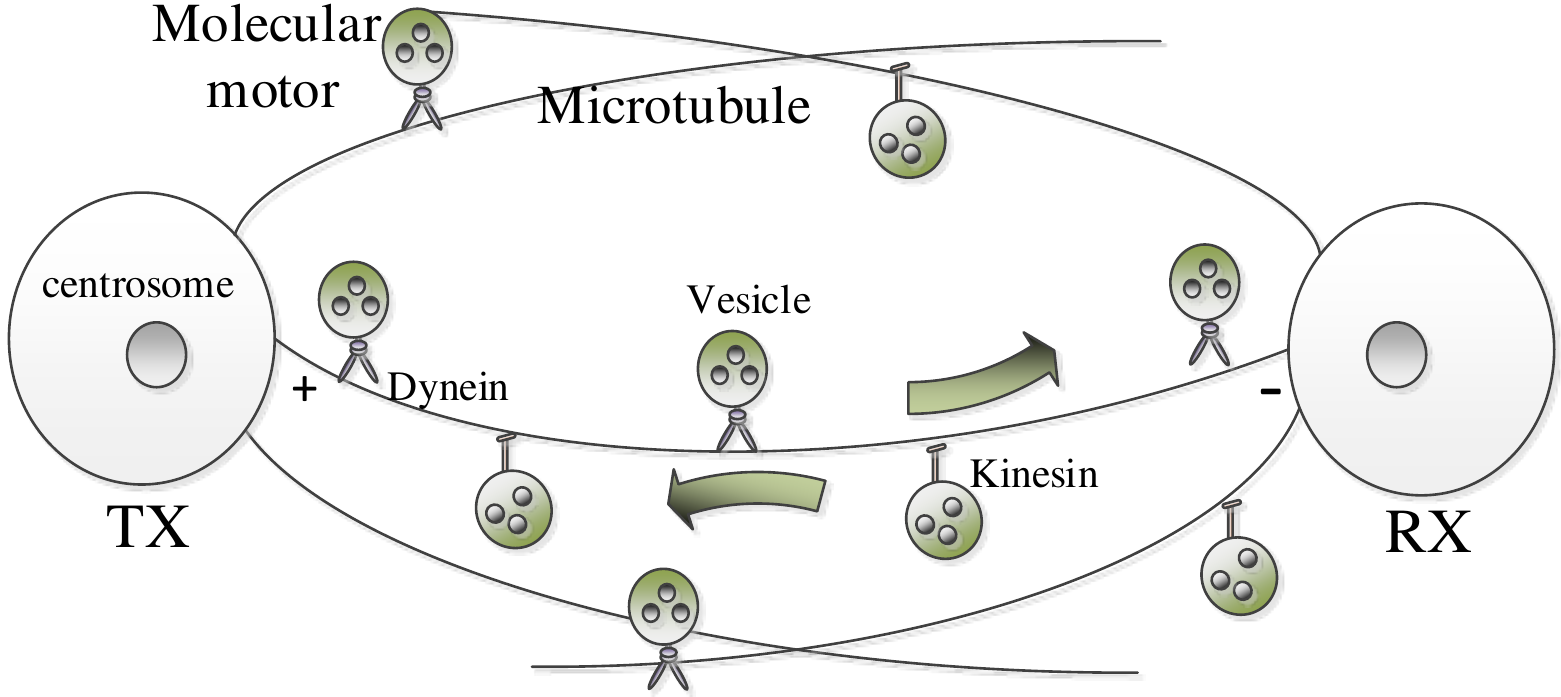}
 \vspace*{-10pt} \caption{Schematic diagram of the molecular motor based MC transmission.} \vspace*{-10pt} \label{3}
\end{figure}

$b)$ Bacterial transport mechanism. The bacteria moves instinctively towards the high chemical concentrations based on the concentration gradient in the environment, namely, the bacteria chemotaxis. As a result, the bacteria can be used as the information carrier to realize MC, in which the messages are encoding into the DNA of plasmid in bacteria {\cite{b10}},{\cite{b11}}. In bacteria based MC, the receiver will emit guiding molecules (e.g., pheromone or nutrients) in the environment to attract bacteria and the message can be exchanged among bacteria through the conjugation process, which is a way to perform the relay function. Fig. 4 is a schematic diagram of bacteria based MC.

\begin{figure}[H]
 \centering
 \includegraphics[width=0.85\textwidth]{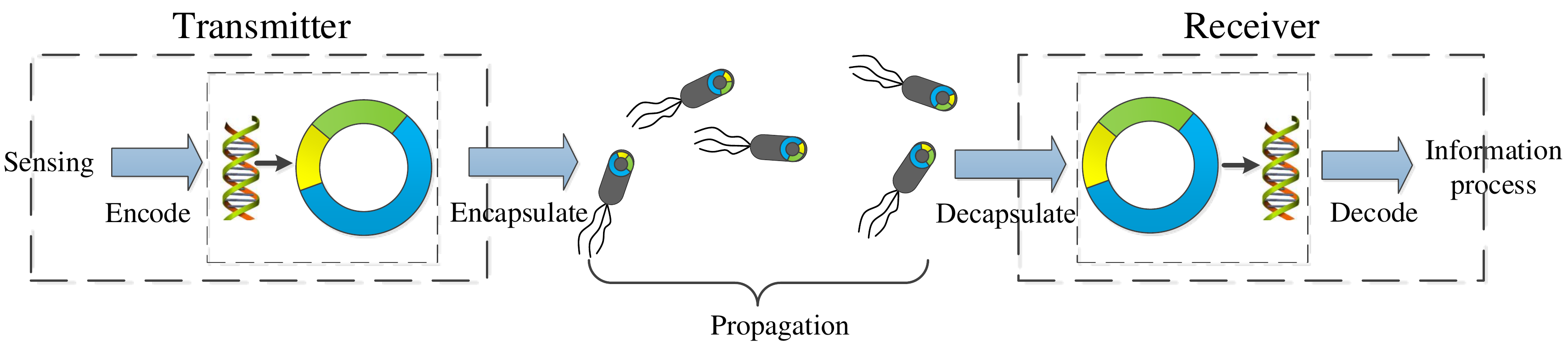}
 \vspace*{-10pt} \caption{Schematic diagram of bacteria based MC transmission.} \vspace*{-10pt} \label{4}
\end{figure}

\subsubsection{Passive Transport Mechanism}

The passive transport is suit to communication in unpredictable, dynamic, or no-infrastructure environments. However, in passive transport, information molecules randomly diffuse in all directions, making it need a large number of molecules to realize communication and a long time to reach a destination increasing with the square of the distance.

$a)$ Free diffusion transport mechanism. The free diffusion is a common molecular transport mechanism intracellulars or intercelluars. The free-diffusion based MC in nature is a kind of radio communication mode that makes the information to be moving in any direction, without the need of additional energy source and communication facilities. Free-diffusion is the process of moving from the high concentration area to the low concentration region in the absence of chemical reaction, and finally the concentration of the state is reached uniformly. The schematic diagram of the free-diffusion based MC transmission is shown in Fig. 5.
\begin{figure}[H]
 \centering
 \includegraphics[width=0.75\textwidth]{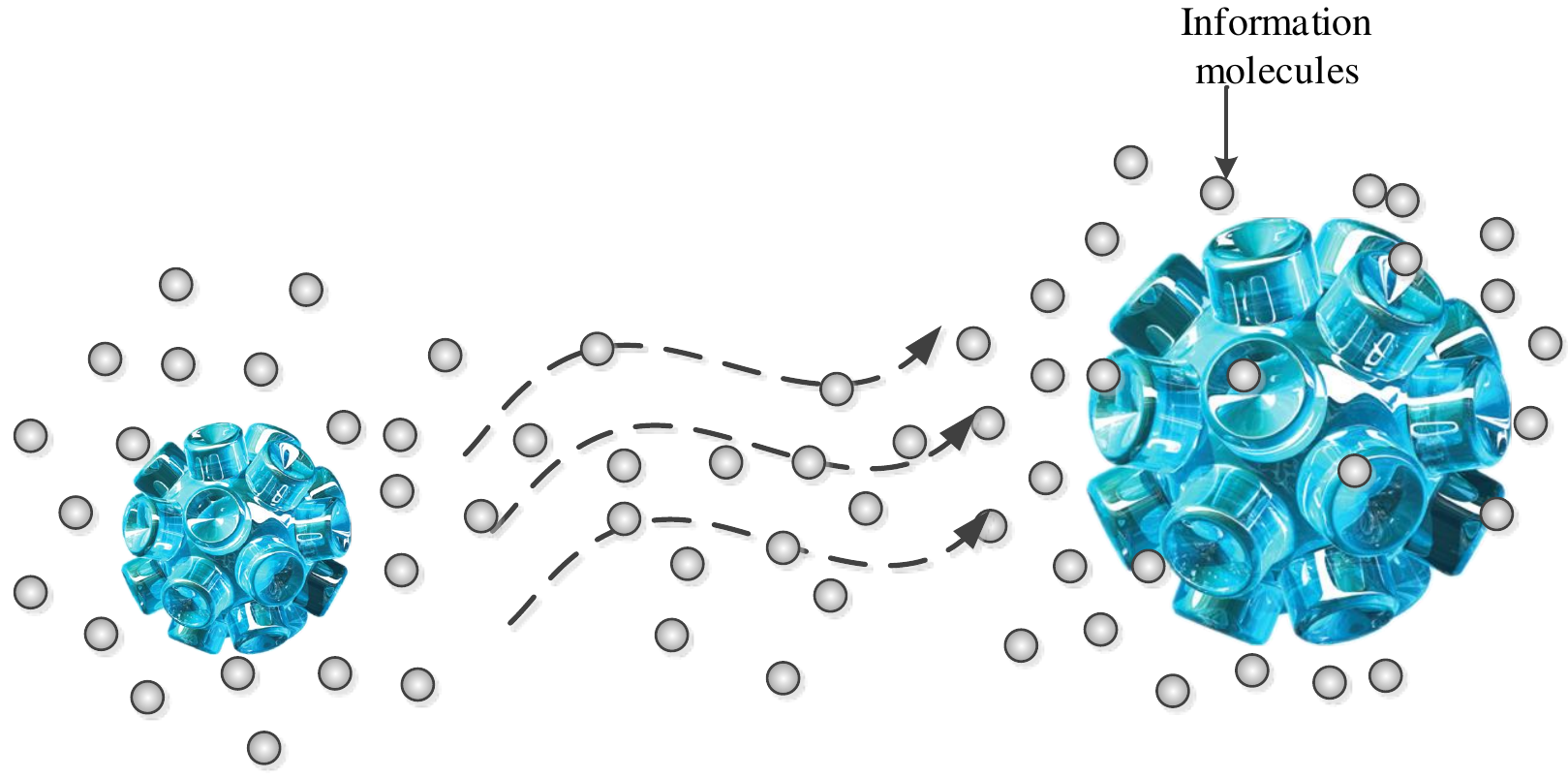}
 \vspace*{-10pt} \caption{Schematic diagram of the free-diffusion based MC transmission.} \vspace*{-10pt} \label{5}
\end{figure}

$b)$ Reaction-diffusion transport mechanism. The reaction-diffusion is a diffusion process that can amplify the molecular signal by chemical reaction (increase the number of the transmitted information molecules). The transmission delay of the reaction diffusion channel can be reduced with the increase of the number of molecules released by the amplifier or the number of amplifiers. Fig. 6 is the schematic diagram of the reaction-diffusion based MC transmission.
\begin{figure}[H]
 \centering
 \includegraphics[width=0.8\textwidth]{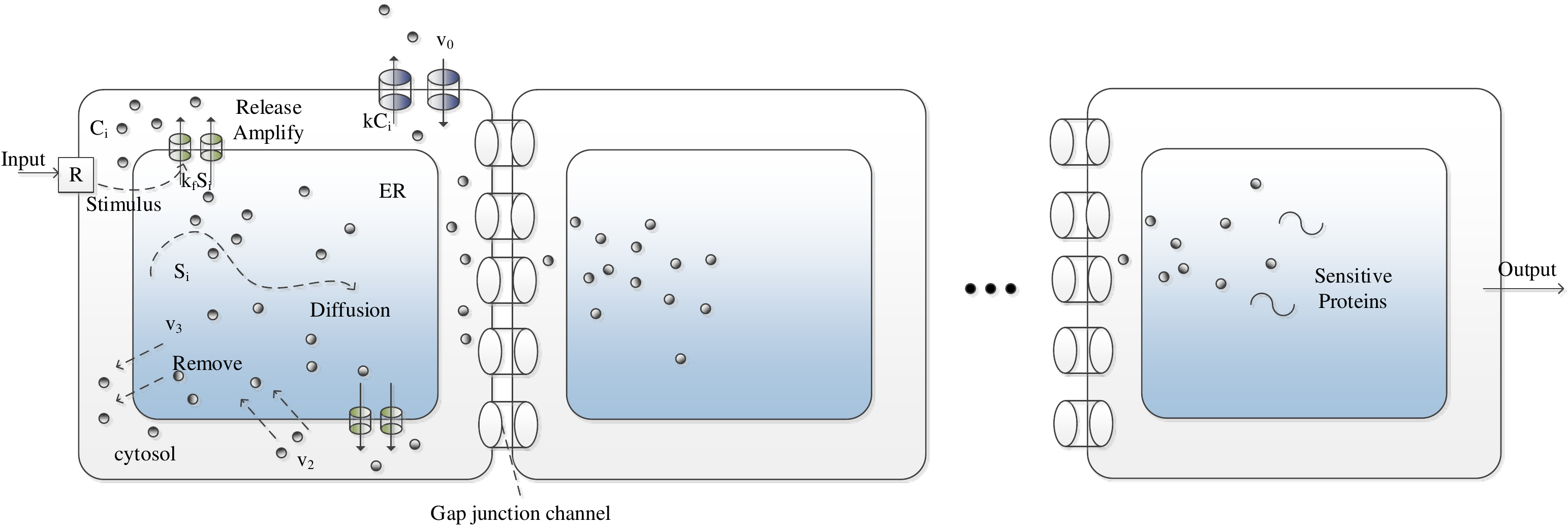}
 \vspace*{-10pt} \caption{Schematic diagram of the reaction-diffusion based MC transmission.} \vspace*{-10pt} \label{6}
\end{figure}

$c)$ Drift diffusion transport mechanism. The drift diffusion is a process of continuous transmission of information molecules with directional drift. The characteristic of the drift diffusion channel needs to consider the rate factor of the fluid medium. The related researches demonstrate that drift diffusion is an effective way for long distance transmission, and the transmission delay and jitter decrease with the increase of drift rate.

Table II lists the characteristics of both passive and active transport transmission mechanisms.

\begin{table*}[tbp]
\centering
\caption{Characteristics of passive transport and active transport mechanism}
\begin{tabular}{|p{5cm}| p{5cm}| p{5cm}|}
\hline
Communication Features &Molecular Communication &Traditional Communication\\ \hline
Transmission Characteristics &Passive transmission &Active transmission\\\hline
The Movement Form of Information Molecule &Diffusion &Directional movement\\\hline
Motion Mechanism of Information Molecule &No chemical energy is driven and influenced by environmental factors &Chemical energy drive\\\hline
Average Transmission Delay &High &Low\\\hline
Demand for Information Molecules &A lot &A few\\\hline
Communication Facilities &No need &Transport molecules, guiding molecules and interface molecules\\\hline
Application Environment &Highly dynamic, unpredictable or lack of communication facilities &Large signal molecule\\ \hline
\end{tabular}
\end{table*}

\section{Characteristic of Diffusion Based MC}

The diffusion based MC is a common form, and its system model is shown in Fig. 7. The communication channel is noisy due to the stochastic nature of diffusion and chemical reactions, and this has significant influences on the diffusion based communication performance. Therefore, a crucial important research areas about the diffusion based MC systems are to analyze and characterize their performances, which mainly consists of the modeling of MC and the system performance analysis, i.e., the channel capacity and error probability.

\begin{figure}[H]
 \centering
 \includegraphics[width=0.75\textwidth]{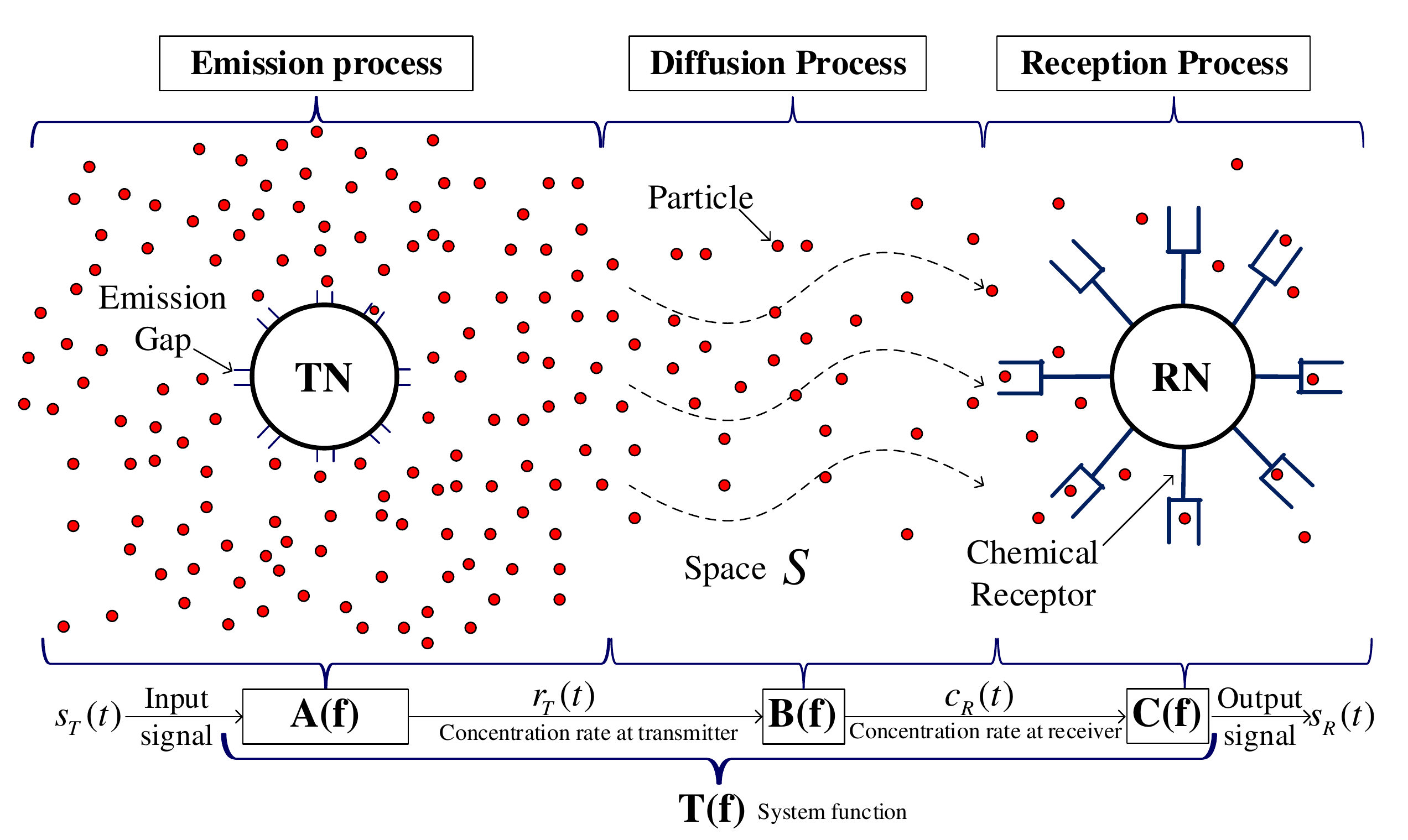}
 \vspace*{-10pt} \caption{The diffusion based MC model.} \vspace*{-10pt} \label{7}
\end{figure}

\subsection{System Model of MC}

There are mainly three different diffusion based MC architectures, which have been modeled on the basis of the encoding type with the information molecules.

The first architecture is modeled in diffusion channel, where the information is encoded and decoded into the molecular emission time and reception time {\cite{b12}}. In order to achieve the precise time instants of the molecular emission and reception, the ideal transmitter and receiver are assumed in this model. In terms of simulation results, the diffusion based MC has poor capacity due to the high freedom of molecular diffusion.

The second architecture is modeled based on the molecular type encoding {\cite{b13}}, where each molecule carries a piece of information according to its molecular composition.

The general architecture is modeled based on the molecular concentration encoding. In {\cite{b14}}, an end-to-end model for free-diffusion based MC is presented, which includes the modeling of the transmitter, the channel and the receiver by the use of electrical parallel RC circuit. Besides the end-to-end normalized gain and delay function are derived for continuous molecular diffusion channels with ligand receptors for detection, meanwhile, in terms of the results, the frequency spectrum considered in the diffusion based MC system ranges from 0 Hz to 1 kHz.

One of the special architectures is modeled based on the reaction-diffusion master equation (RDME), which is a well known model with both diffusion and chemical reactions {\cite{b15}}. This model is called reaction-diffusion master equation with exogenous input (RDMEX), which considers the transmitters as time series of signaling molecule counts, while diffusion in the medium and chemical reactions at the receivers are modeled by RDME. By the use of Markov model, the corresponding Markov state, which means the number of the information molecules in discrete location, changes according to the RDMEX during each discrete time slot. Another stochastic model is investigated using birth-and-death processes in diffusion based MC with or without flow {\cite{b16}}. Three environmental effects in this model are considered, namely, particle absorption, particle generation and spontaneous emission phenomena.

\subsection{Performance Analysis and Influence Factors for MC}

Although MC has been presented in nature for billions of years, it was only recently that engineering MC systems has been proposed. Therefore, compared to traditional electromagnetic based communication systems, MC is still in its infancy. Information theory is the mathematical foundation of any communication system. Since the Shannon theorem is not applicable for MC, the general capacity of the MC channel with memory and noise, over all possible modulation schemes, is an important open problem.

In{\cite{b17}}, the information theoretic models for diffusion based molecular nano-communication model which includes a nano-transmitter and a nano-receiver is presented, where the information rates based on simulation between channel inputs and outputs are analyzed when the inputs are independent and identically distributed (i.i.d). In{\cite{b18}}, the authors investigate the channel capacity for the ligand-reception model, where the information bits are encoded in the time variation of the concentration of molecules. In {\cite{b19}}, the MC transmission of binary coding is modeled, where the effect of the memory channel is studied, and the corresponding transmission capacity is given. It is a pity that the influence of the molecular noise is not considered. The MC channel capacity in a one dimensional environment is analyzed in {\cite{b20}}, which demonstrates the channel capacity is largely affected by the life expectancy of molecules. In {\cite{b21}}, a closed-form expression for the free-diffusion based MC channel capacity is proposed. The capacity expression is independent from any coding scheme and takes into account the two main effects of the diffusion channel: the channel memory and the molecular noise.

The channel capacity of ligand-based MC system with continuous molecule emission scheme is studied in {\cite{b22}}, meanwhile, taking account of the influence of channel memory factors, the lifetime of the molecules can not be neglected as an important factor. The MC is applied to the particulate drug delivery systems (PDDS) for the first time in {\cite{b23}}. At the same time, the paper points out that the PDDR capacity is derived analytically considering all related noise effects and the constraints on the drug injection. Similar to {\cite{b23}}, the paper {\cite{b24}} provides some analysis on the effect of human temperature variations for MC. The results of simulation indicate that the molecular concentration decreases as the system temperature increases, while the capacity increases as the system temperature increases. In {\cite{b25}}, the authors investigate a bacterial point-to-point communication system with one transmitter and one receiver, where capacity is given by the numerical algorithms based on the ligand-receptor binding process. At the same time, the authors derive an upper bound on the capacity of the ligand-receptor for a binomial channel model and a lower bound when the environment noise is negligible. In {\cite{b26}}, the authors also acquire the lower and upper bounds on capacity of channel. The difference to {\cite{b25}} is the information is modulated on the releasing time of information particles, and decoded from the time of arrival at the receiver.

Thus, the channel models with different parameters lead to different channel capacity. Because the mathematical models of the molecular diffusion channel with the molecular noise are not uniform, the influence of the related MC transmission capacity is not fully revealed.

Information molecules in the transmission process may experience a variety of non-deterministic factors influence, such as collisions among molecules, chemical reaction in the environment, thermal noise, et. One of the main challenges in diffusion based MC is the characterization of the noise. Two noise sources are firstly presented in {\cite{b27}}, namely, the particle sampling noise and the particle counting noise, which are related to the transmitter emission and the signal propagation in the diffusion channel respectively. The noise sources are modeled both in the physical and stochastic fashion. The physical model provides a mathematical analysis of the physical processes about the noise generation, and provides a way to simulate the generation of noise in MC. While the stochastic model characterizes the noise generation using random processes and their associated parameters. The same authors present a reception noise source due to the ligand-receptor binding at the receiver in diffusion based MC {\cite{b28}}. Similar to {\cite{b27}}, the ligand-receptor binding is modeled through two different approaches, namely, the ligand-receptor kinetics and the stochastic chemical kinetics, which provide the mathematical expressions for the noise source simulation and closed-form solutions for the noise stochastic modeling respectively. In {\cite{b29}}, a mathematical model for molecular noise resulting from the random nature of the Langevin force is proposed, which is the first study of molecular noise from a microscopic perspective. In the case of the diffusion based MC with information conveyed in the releasing time of molecules, the system noise is modeled as the inverse Gaussian (IG) distribution in {\cite{b30}}, which results in the additive inverse Gaussian noise (AIGN) channel model.

MC channels have memory since the partial information molecules arrive in later time slots which results in the generation of ISI.
In general, the ISI is the overlap between the current received signals and the previous received signals in molecule concentration which are transmitted from a single molecular transmitter. If the system has multiple transmitter, Co-Channel Interference (CCI) will also be considered in the MC system, which is identified as the overlap between received signal transmitted by a single transmitter and all the received signals transmitted by the other concurrent transmitters.

In {\cite{b31}}, the first effort to characterize the ISI is referred to a unicast MC system with the most popular modulation technique, i.e., binary amplitude modulation. The authors determines that the ISI is influenced by three main
factors, i.e., the distance between transmitter and receiver, the ``on" time of a bit in on and off keying (OOK) modulation, and the data rate of transmitted bits.

In {\cite{b32}}, the ISI is considered under another specific modulation technique which the information is encoded with the order of the transmitted molecules. Meanwhile, the effects of CCI in diffusion based MC are investigated in {\cite{b33}}, which is related to the transmission range, radius of the nano-machines and the distance between the transmitters.

The joint effects and closed-form expression of both the ISI and the CCI are analyzed in {\cite{b34}}, which involves two specific modulation schemes, namely the baseband modulation and the diffusion wave modulation. It is determined that the diffusion wave modulation scheme shows lower interference values, and the higher modulating frequency has lower effects of the ISI and the CCI on the communication channel.

A statistical-physical modeling of the interference in molecular nano-networks based on diffusion is presented in {\cite{b35}}, where multiple transmitters are emitting molecules simultaneously, which is a general model independent from specific coding and modulation techniques. Through assuming the received molecular signal follows a stationary Gaussian Process (GP), an analytical expression of the log-characteristic function is achieved, which leads to the estimation of the received Power Spectral Density (PSD) probability distribution. The results indicate that the probability of interference has very low values for lower frequencies and distance range.

A unifying model for noise, CCI, and ISI is presented in {\cite{b36}}, where a general time-varying expression is presented for the expected impact of a noise source that is emitting continuously. Meanwhile, the interference can be approximated as a noise source.

\section{Coding and Modulation Techniques for MC}

For MC, the information coding and modulation techniques can be summarized into the following five aspects, as shown in Fig. 8.

\begin{figure}[H]
 \centering
 \includegraphics[width=0.75\textwidth]{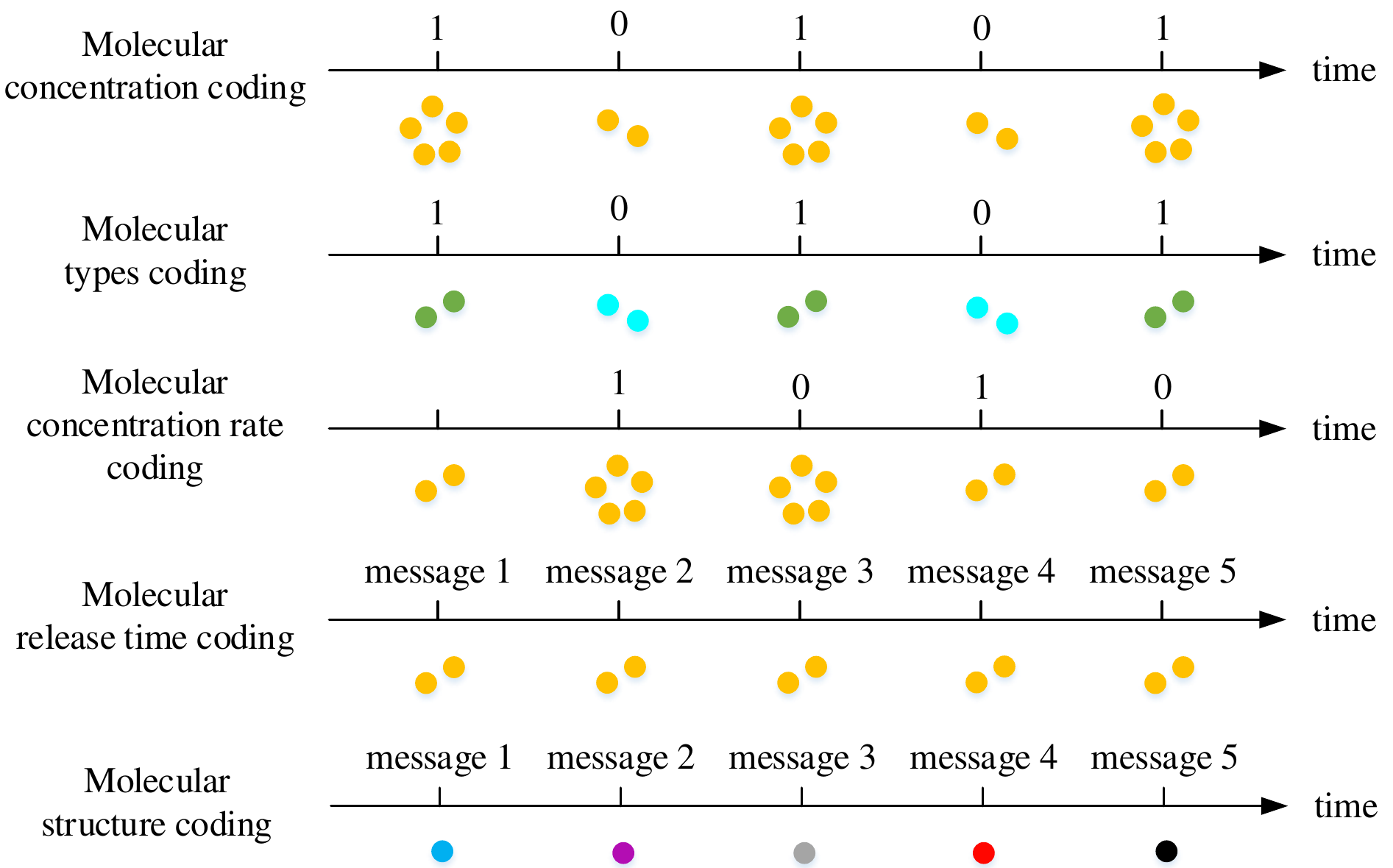}
 \vspace*{-10pt} \caption{Coding and modulation techniques in MC. } \vspace*{-10pt} \label{8}
\end{figure}

$1)$ Molecular concentration coding. This coding technique is parallel to amplitude modulation (AM) technique {\cite{b37}},{\cite{b38}} in traditional communication, using information molecules of different concentration on behalf of bit $0$ and $1$ {\cite{b13}},{\cite{b39}}. The receiver detects the concentration of information molecules received, which can be used to acquire messages by comparing with the predetermined threshold {\cite{b40}}.

$2)$ Molecular types coding. This coding technique uses different types of molecules to represent different bits of information. Receiver determines the information based on the reaction type between molecules and the corresponding receptor (ligand-receptor binding).

$3)$ Molecular concentration rate coding. This coding technique is parallel to the frequency modulation (FM) technique {\cite{b41}} in traditional communication, where the molecular concentration rate (the time of the concentration $dC(t)/dt$) is used to represent different symbols {\cite{b27}},{\cite{b28}}.

$4)$ Molecular release time coding. This coding technique is parallel to the phase modulation (PM) technique {\cite{b42}} in traditional communication, where the information is encoded by the releasing time of information molecules {\cite{b30}},{\cite{b43}}. The precondition is that the nano-machines need to have precise time synchronization mechanism {\cite{b44}}.

$5)$ Molecular structure coding. Unlike the binary coding mode widely used in computer communication system, the molecular structure coding is a quaternary information coding technique {\cite{b45}},{\cite{b10}} based on the sequence of DNA nucleotides (adenine, thymine, cytosine and guanine). This coding technique improves the throughput of the channel by carrying a large amount of information, and has good noise immunity {\cite{b46}}.

Particularly, a special coding and modulation technique called depleted MoSK (D-MoSK) is proposed in {\cite{b47}}, where molecules are released if the information bit is $1$ and no molecule is released for $0$. This technique requires the reduced number of the type of molecules for encoding.

Coding and modulation techniques are mainly used in reducing the ISI and noise. In {\cite{b48}}, the authors propose an ISI-free coding technique to increase the communication reliability while keeping the encoding/decoding complexity reasonably low to mitigate ISI. For a pair of molecules, the molecules that has been sent earlier may arrive later than the other one. This phenomenon is a crossover. A level-$l$ ISI-free code guarantees that no matter how many crossovers happen to the encoded sequence, the decoded information is error-free if all the crossovers are no more than level-$l$. Two types of molecules are used to code to reduce ISI in {\cite{b49}}. For example, the transmitter uses type-$A_{1}$ molecules in odd time slots and type-$A_{2}$ in even time slots. As the molecule types are different in two subsequent time slots, the ISI is significantly reduced. Similar to {\cite{b49}}, the method based on the order of information particles released by the transmitter to reduce ISI is presented in {\cite{b32}}. Bit-$0$ is encoded by releasing information particle $a$ followed by information particle $b$, and bit-$1$ is encoded by release of $b$ followed by $a$. Binary version of one modulation technique utilizes two types of information particles $A$ and $B$. The modulator decides which molecule to send depending on the current bit and previously sent transmission bit. Using this technique, the ISI can be reduced in {\cite{b34}}.

A major concern has always been on the reliability of the data at the receiver and the Error Correction Codes (ECCs) is a promising technique to control or correct any errors introduced. In {\cite{b50}},{\cite{b51}}, the uses of Hamming block codes are presented to increase transmission reliability and lower transmission power. In terms of the energy budget, the results indicate that coding is beneficial for distances above a few tens of meters while the shorter ranging networks are not suitable. Therefore, new codes should be investigated within the context of the limited energy availability when the range of energy effective operation is to be enhanced, particularly at high BER values. In {\cite{b52}}, two ECCs, namely, Euclidean geometry low density parity check (EG-LDPC) and cyclic Reed-Muller (C-RM) codes are considered within a MC system, where these codes are compared against the Hamming code with respect to both coding gain and energy cost. According to the results, the designer can determine which code should be employed by the critical distance (the distance at which the energy gain overcomes its operational energy cost) and the energy cost for different operating BERs.

\section{Multi-hop Nano-Networks Based on MC}

The nano-network based on diffusion MC is composed of a set of nano-machines located at both ends of the fluid medium, in which the information communication is carried out by means of diffusion. The process of diffusion MC mainly consists of three main stages: transmission, propagation, and reception. First, the transmitter nano-machine codes information through the mode of molecular releasing and the transmission of the molecule causes a local concentration changes. Then, information spread in the medium through the diffusion process. Finally, the receiver can decode the transmitted information by estimating the molecular concentration near it.

\subsection{Nano-Networking Model}

Due to the extreme simple of the nano-machines and the diffusion characteristic that the propagation time increases with the square of the distance, using a single transmitter may be impractical when the receiver is far away. One method to execute the remote communication is the participation of the relay. Fig. 9 shows the scenario of the nano-network based on the relay MC.

\begin{figure}[!t]
 \centering
 \includegraphics[width=0.75\textwidth]{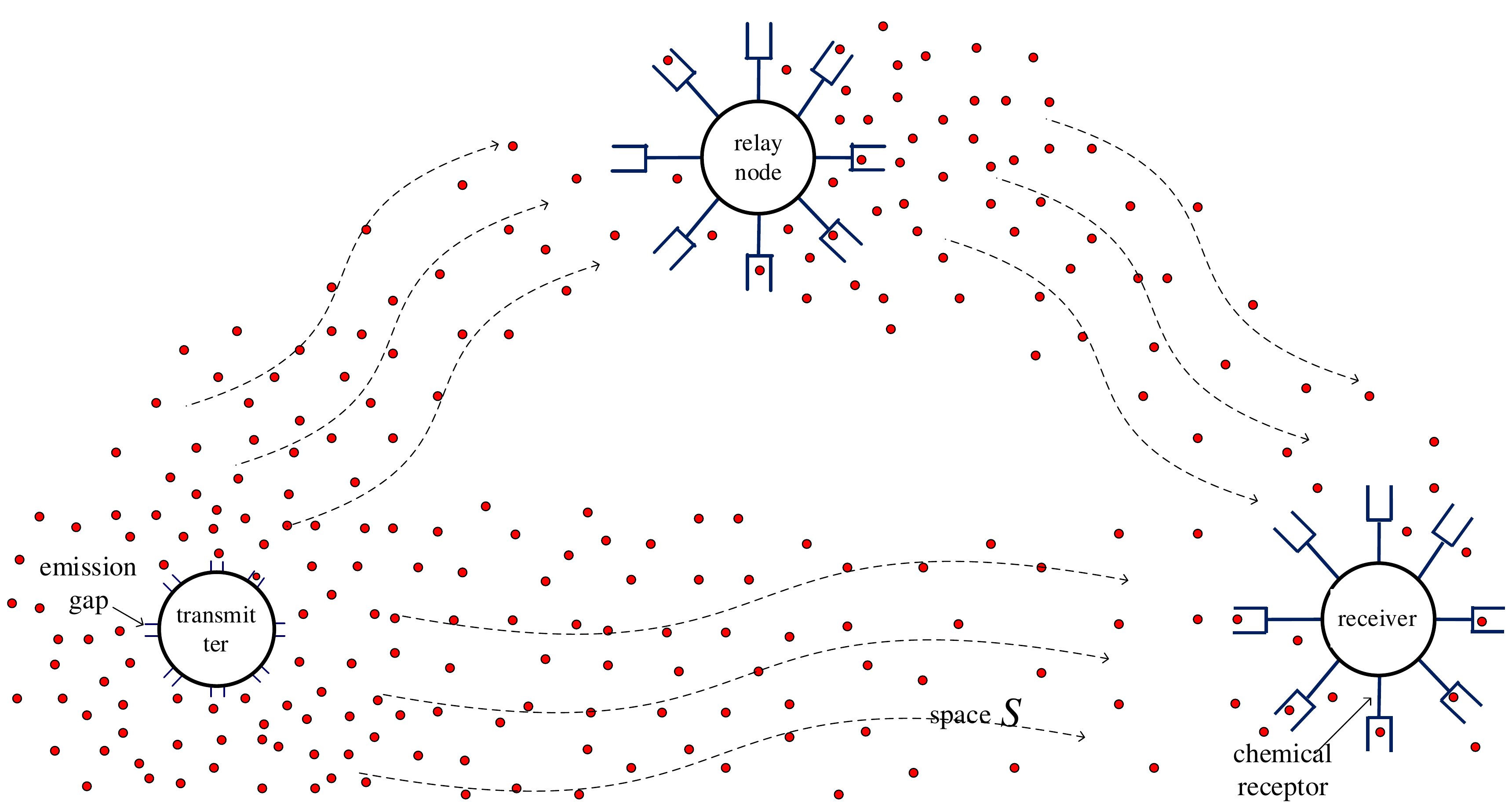}
 \vspace*{-10pt} \caption{The relay based transmission model for MC. } \vspace*{-10pt} \label{9}
\end{figure}

The research of relay channel is still in the embryonic stage, and the related research literatures are hard to be found. From the information theory point, a relay channel model that explores communication capacity is introduced in {\cite{b53}}. It also reveals the relationship between the relay channel characteristics and the channel capacity.

In {\cite{b54}}, both the amplify-and-forward (AF) and decode-and-forward (DF) relaying protocol are investigated in diffusion based MC with bio-nodes consisting a population of bacteria and M-ary modulation scheme, where the relay with DF scheme first decodes the received message, and then re-encodes the detected message for re-transmission, while the relay with AF scheme simply amplifies the received message for re-transmission. Single-molecule-type relaying and two-molecule-type relaying are considered respectively in AF scheme, where the capacity gain is more significant for the low range of concentrations at the receiver with the single-molecule-type relaying, while the two-molecule-type relaying has the great gain for both the ranges of low and high concentrations at the receiver. For the DF relaying scheme, the maximum a posteriori (MAP) detection is used, which decreases the error probability but becomes invalid for high ranges of concentration at the receiver.
In {\cite{b55}}, the relay scheme is also adopted to solve the problem that the communication reliability decreases with the increasing of distance between the transmitter and the receiver. The authors use repeated coding to make the system have a low complexity, and the simulation results show that error rate of the system is greatly reduced.
A more detailed research for the DF relaying scheme by the same author and with the same physical model as well as the modulation scheme appears in {\cite{b56}}, where a generalized form of Maximum Ratio Combining (MRC) is used to find the optimal relaying in terms of the error probability.

Both the aforementioned efforts are investigated in an MC system with steady-state concentrations which is unpractical for a continuous infinitely-lasting molecule emission. Conversely, the first publication on relay systems with the time-dependent molecular concentrations can be found in {\cite{b57}}, where a two-hop MC nano-network based on diffusion is investigated for the DF relaying protocol with two relaying schemes, namely, two-molecule-type relaying and single-molecule-type relaying considered. Two-molecule-type relaying means that two different types of information molecules are used to transmit and detect at the relay node, while only one type of molecule is applied in the single-molecule-type relaying and the self-interference is introduced. There are two working modes in relay node, i.e., full-duplex and half-duplex, where the half-duplex is a method to mitigate self-interference. However, the adaptation of the decision threshold is an effective mechanism to mitigate self-interference for full-duplex mode.

In {\cite{b58}}, a DF relaying scheme is introduced in a diffusive MC system with the time-dependent MC encoding influenced by both noise and channel memory, where the impact of emission process and the optimal relay positions in terms of the bit error rate (BER) are considered and investigated for the first time. The same authors with {\cite{b57}} consider a multi-hop MC nano-networks with multiple relays {\cite{b59}}, which is a expanding work for {\cite{b57}}. A new relaying scheme with different types of information molecules is utilized at all relay nodes and the backward inter-symbol interference (backward-ISI), and forward-ISI is firstly identified for the two-molecule-type relaying and single-molecule-type relaying schemes. Furthermore, the optimal number of molecules released by the nano-transmitter and the optimal detection threshold of the nano-receiver for minimization of the expected error probability of each hop are investigated. Meanwhile, the same authors utilize AF relay protocol with fixed and variable amplification factor in a two-hop MC nano-networks based on diffusion{\cite{b60}}, where a closed-form expression for the optimal amplification factor at the relay node for minimizing the expected error probability of the network is derived.

\subsection{Multi-hop Nano-Networking Technology}

Network coding is an advanced networking technology, in which the relay is used not only as a conveyor, but also as an encoder to improve system throughput and reliability. Two main classes of network coding techniques in traditional wireless networks are straightforward network coding (SNC)
and physical layer network coding (PNC) {\cite{b61}}, where the transceivers send their messages in two time slots to the relay and the relay sends the bitwise XOR of the messages in the next time slot in SNC, while the transceivers simultaneously send their messages in one time slot in PNC.

In {\cite{b62}},{\cite{b63}}, an XOR logic gate at the relay node is used to execute the network coding, which is parallel to the SNC scheme in traditional wireless networks. The rate-delay tradeoff with network coding in molecular nano-networks is investigated in {\cite{b62}}, where the results indicate that the coded case has higher rate with the same delay or lower delay with the same rate compared to the uncoded case. In {\cite{b63}}, the closed-form expressions for error performance of network coding in molecular nano-networks based on diffusion is firstly derived by adopting pulse based modulation with energy detection.

\subsubsection{Network Coding}

Compared to the SNC scheme, the PNC scheme decreases the requirement of time slots number for two-way communications and has a lower BER. PNC is practicable in traditional wireless networks due to the characteristic that the superimposed EM waves can be decomposed through the signals canceling out at the transmitters. However, in MC, the transmitted signals cannot be negated naturally. M. F. Ghazani \emph{et al.} utilize the molecular reaction to cancel out the signals in {\cite{b64}}, where another network coding scheme parallel to the PNC is presented with the ligand-receptor binding process. Specifically, different types of molecules are used at transmitters respectively and can react with each others, where the reaction rate is much more rapid than the molecule binding rate to the relay node. As a result that almost no molecule binds to the relay with both messages ``1" at the transmitters and only one type molecule binds to the relay with one message ``1" and one message ``0" at the transmitters. In the first case, the relay receives no molecule, while the relay sends a third type molecule in the next slot.

\subsubsection{Medium Access and Routing}

In {\cite{b65}}, the communication performance of a molecular multiple access channel is investigated, which reveals that different nano-machines must access the medium by using different kinds of molecules to carry message, while transmission scheduling is impractical due to low-end computation and synchronization capabilities of nano-machines. Similarly, routing protocols require minimal overhead and computational complexity due to the very low-level memory and processing capabilities of nano-machines. An opportunistic routing (OR) protocol based on distance and concentration gradient information is proposed in {\cite{b66}}, where the addressing problem is solved by the molecular broadcast. A routing mechanism by selecting the optimal releasing point for emitting information molecules to reduce the communication distance is investigated in {\cite{b67}}, where the molecular hit probability can get a great increase in the case of lower average propagation delays.

\subsubsection{Interconnection}

Until now, many different kinds of MC have been presented, and they are suitable for different locations within the body area. It is a opportunity to interconnect all these heterogeneous MC nano-networks to realize more complex tasks in the entire body area. A solution may be the biological circuits within the artificial cells, which can provide a set of genetic instructions that mimic the classical gateways to perform the transformation among different types of information molecules{\cite{b68}}. Since the MC is used in the body area, another challenge is to realize the interconnection between the intra-body and inter-body devices by encapsulating biological nano-sensors and electromagnetic (EM) nano-communication units within the artificial cells, which can be competent for interfacing between the chemical and electrical domains. In the artificial cells, biological nano-sensors is responsible for the reception of the information molecules, while the EM nano-communication units is responsible for the communication with the device in vitro through EM. The bio-compatibility and the ability to produce sufficient power for EM emission are both the emergency challenge for the nano-network interconnection.

\subsection{Applications}

Particle drug delivery system through the nano-particle drug targeted delivery to the specific lesion site, bypass the physiological barriers and will not affect other healthy parts, avoiding the interference of external factors. As shown in Fig. 10, the transmission process of drug particles in the body can be considered as a MC system.

\begin{figure}[!t]
 \centering
 \includegraphics[width=0.8\textwidth]{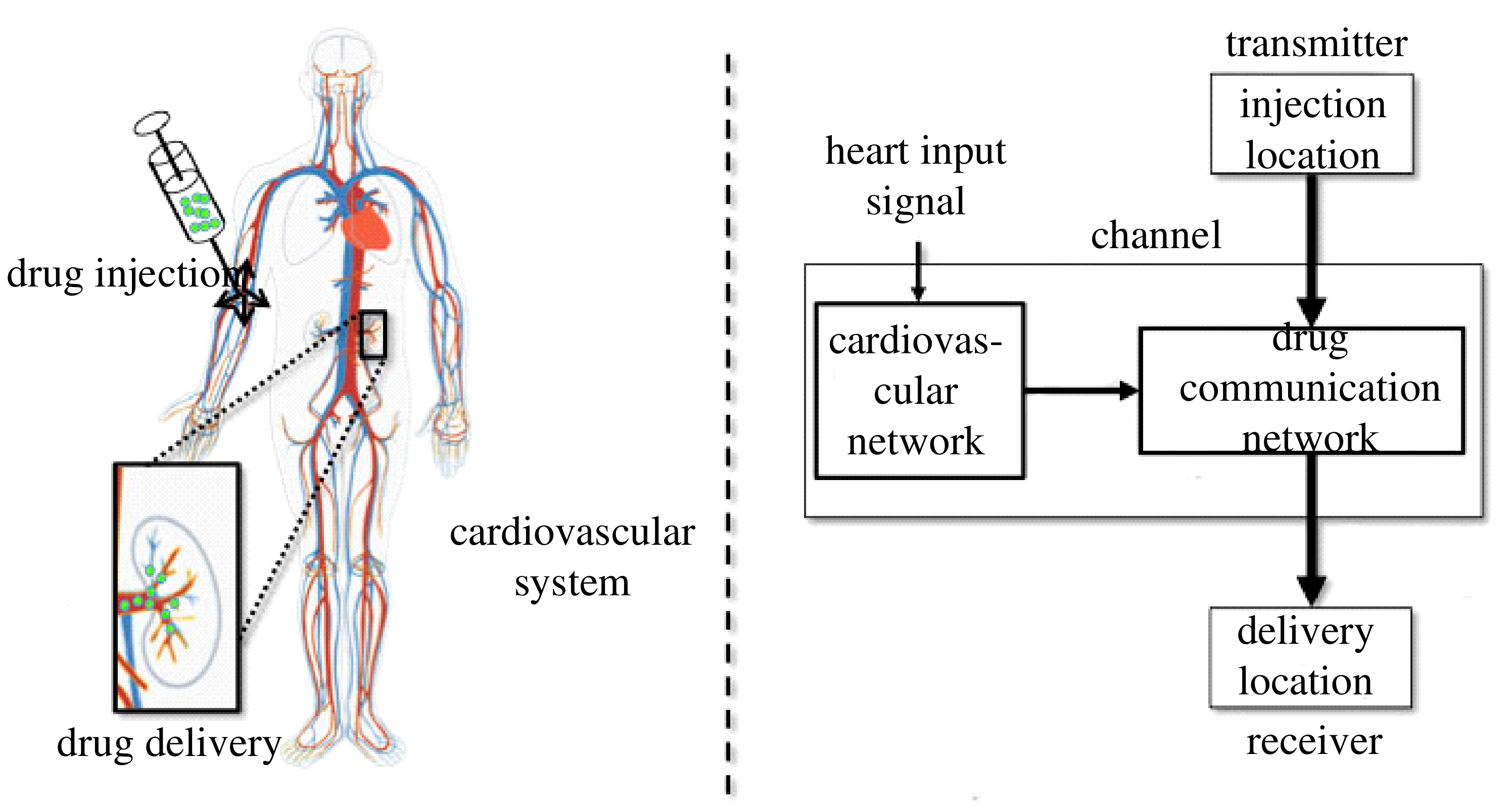}
 \vspace*{-10pt} \caption{Particulate drug delivery and analysis models. } \vspace*{-10pt} \label{10}
\end{figure}

The MC theory is applied to the particulate drug delivery systems, abstraction of drug information molecules from the transmitter to the receiver, and through communication engineering theory tools to realize the control and prediction of particulate drug delivery{\cite{b69}}. The drug injection site is the transmitter, the receiving end is the drug delivery site, and the propagation process of the particle drug is abstracted into the MC channel model. The MC channel model is divided into two parts: the cardiovascular network model and the drug transmission network model. Cardiovascular network model calculates the distribution of blood flow in the cardiovascular system by analyzing the distribution of blood flow and blood pressure. The drug delivery network model allows the analysis of the drug injection rate and the drug injection location distribution to obtain the analytical expression of drug delivery rate at target site. In the process of propagation, the drug particle is influenced by many kinds of noise, such as injection noise, transmission noise, ligand-receptor noise, et. Drug particles propagation in the cardiovascular system by the influence of the noise is simulated by Monte Carlo. The results show that the analysis model and simulation results have good consistency, verifying the model for MC can be conveniently used for particulate drug delivery system analysis and optimization {\cite{b23}}.

MC is not only used in the delivery of drug particles, but also has a wide range of applications in the activation of the immune system, tissue repair, nano surgery and so on. In {\cite{b70}}, molecules transport through artificial nano-machines, and trigger the immune response, enhancing their ability to fight against a specific threat. In {\cite{b71}}, the authors analyze a mechanism by which the humoral immune response of T-lymphocytes is activated. These T-lymphocytes produce a positive feedback amplification signal that is passed to the B-lymphocytes, thereby releasing the antibodies. At present, some techniques for tissue repair have been studied, and the potential for organ reconstruction may be achieved {\cite{b72}}. Thereby, the patients with organ failure can achieve an alternative treatment in the case of the actual organ shortage. An application in this area is to detect the biological function of cells, issues and organs by monitoring the behavior of calcium ions. In fact, the short distance communication through the calcium ion signal is the study of a large spread of communication, where information is usually encoded by modulation of the ion concentration {\cite{b73}}. In recent years, research has focused on communications between nano-machines that may interface with tissues within organs by means of calcium ions {\cite{b74}}{\cite{b75}}. In {\cite{b76}}, the authors analyze the effect of tissue deformation on the propagation of calcium ions, and the reliability of the transmission depends on the physical state of the structure. A set of coordinated nano-machines, equipped with a specific execution tool, is envisioned to be installed inside the patient's body to perform the nano surgery {\cite{b77}}. In {\cite{b78}}, the authors have proposed an atomic force microscopy based nano-robot with integrated imaging, manipulation, analysis and tracking functions, important for cellular-level surgery on live samples.

\section{Open Issues and Challenges}

The bio-inspired based MC is a promising short distance communication technology, providing efficient and reliable information exchange for nano-machines, and supporting on the network to perform more complex tasks. At present, the researches of MC and related fields are in the initial stage, in which problems and challenges are mainly divided into two aspects, namely, theoretical research and practical application.

\subsection{Channel Model for MC}

Although a mass of channel modelings have been presented and investigated for MC, the practical models are still urgent in order to acquire the accurate results and show the performance bounds for special applications. In particular, the advanced channel model for MC includes as the following:
\begin{itemize}
 \item Finite boundary conditions are needed for the applications within a limited environments.
 \item In terms of short distance applications, spherical nano-machines are preferable to the point source models.
 \item The decay of information particles or the life cycle should be introduced into MC, particularly for application of long distance MC.
 \item Imperfect synchronization is more practical for the MC with the molecular releasing time to code.
 \item Diffusion modeling where the information particles interact with other chemicals should be used in MC within impure fluid medium.
\end{itemize}

In the future, the channel models for MC should be further designed based on the movement of biochemical molecules in different environment. Based on the advanced channel models, the theoretical performance bounds can be presented, which indicates the presentence of key techniques.

Accurately, the research of Shannon＊s information theory for MC is still an open issue since there are no analogy against the signal to noise ratio (SNR) in the traditional communication system. Furthermore, the channel memory and various noise sources make it difficult to derive the channel capacity with closed-form expressions.

\subsection{Self-organizing Nano-Network}

Existing nano-machines do not have the ability of complex information processing, node calculation, and storage capacity. Because more complex tasks need interconnection among nano-machines, it should form a self-organizing nano-network. In the previous literatures, very few works have considered multiuser channels and medium access control technique, which is the basic feature of a nano-network. Furthermore, due to high node density and node random mobility, addressing and routing in the MC based nano-network are key challenges. Since there are a huge number of biochemical molecules in the body area, they should be self-organized, which is beneficial to enhance the robustness. In the future, the self-routing, self-addressing, and self-healing may be the potential solution for MC.

\subsection{Biochemical Application}

The theoretical research challenges aforementioned have been proposed based on channel model, information theory, and self-organizing network. Besides these aspects, the MC practical application is still challenges, which can be concluded as follows.

\begin{enumerate}
 \item The simulation and experimental platform of MC and self-organizing nano-network need to be established, in which the body model, human brain tissue, breast tissue and arterial capillaries are structured based on three-dimension printing technology.
 \item IEEE Standards Association has founded the IEEE P1906.1 project to provide recommended practice for MC and nano-scale networking framework. The further standards need to be urgently established.
\end{enumerate}

\section{Conclusion}

Molecular communication (MC) is a communication breakthrough with a wide range of potential biochemical applications in the future. Although MC is still in its infancy, there have been many advancements in developing theoretical research models and methods. In this paper, the basic principles of diffusion based MC and the corresponding key techniques have been comprehensively surveyed, including the transmit communication mechanism and the corresponding key techniques. The most advanced achievements relative to the diffusion based MC are especially summarized, such as the MC based mathematical analysis modeling, system performance analysis with the inter-symbol interference and noise affect, information coding, and the modulation techniques. Meanwhile, the modeling technique and related application for nano-networking have been discussed as well.

Looking to the future, the diffusion based MC applying in the body area nano-network has a promising application prospects. To promote the applications, the theoretical performance and its influencing factors should be exploited further, and the corresponding communication transmission, signal processing, and self-organizing network should be researched further. Meanwhile, the practical applications need the experimental platforms to demonstrate the feasibility of MC. Note that, as an emerging communication technology, the information security and personal security should be exploited as well.

\section*{Acknowledgment}
This work was supported in part by the National Natural Science Foundation of China (Grant No. 61361166005), the State Major Science and Technology Special Projects (Grant No. 2016ZX03001020006), and the National Program for Support of Top-notch Young Professionals.

\begin{IEEEbiographynophoto}{Jiaxing Wang}
received the M.S. degree in signal and information processing from Yanshan University, Qinhuangdao, China, in 2015. She is currently pursuing the Ph. D degree at the key laboratory of universal wireless communications (Ministry of Education) at BUPT. Her research focuses on molecular communication and body area nano-network.
\end{IEEEbiographynophoto}
\begin{IEEEbiographynophoto}{Bonan Yin}
received the B.S. degree in telecommunication engineering from Jinlin University,China, in 2015, where he is currently working toward the master's degree with the key laboratory of universal wireless communications (Ministry of Education) at BUPT. His research focuses on molecular communication and body area nano-network.
\end{IEEEbiographynophoto}

\begin{IEEEbiographynophoto}{Mugen Peng}
(M'05--SM'11) received the B.E. degree in Electronics Engineering
from Nanjing University of Posts \& Telecommunications, China in
2000 and a PhD degree in Communication and Information System from
the Beijing University of Posts \& Telecommunications (BUPT), China
in 2005. After the PhD graduation, he joined in BUPT, and has become
a full professor with the school of information and communication
engineering in BUPT since Oct. 2012. During 2014, he is also an
academic visiting fellow in Princeton University, USA. He is leading
a research group focusing on wireless transmission and networking
technologies in the Key Laboratory of Universal Wireless
Communications (Ministry of Education) at BUPT, China. His main
research areas include wireless communication theory, radio signal
processing and convex optimizations, with particular interests in
cooperative communication, radio network coding, self-organization
networking, heterogeneous networking, and cloud communication. He
has authored/coauthored over 40 refereed IEEE journal papers and
over 200 conference proceeding papers.

Dr. Peng is currently on the Editorial/Associate Editorial Board of
\emph{IEEE Communications Magazine}, \emph{IEEE Access},
\emph{International Journal of Antennas and Propagation (IJAP)},
and \emph{China Communications}. He has been the guest leading editor
for the special issues in \emph{IEEE Wireless Communications},
\emph{IJAP} and \emph{International Journal of Distributed Sensor
Net- works (IJDSN)}. He received the 2014 IEEE ComSoc AP Outstanding
Young Researcher Award, and the Best Paper Award in IEEE WCNC 2015, WASA 2015, GameNets 2014,
IEEE CIT 2014, ICCTA 2011, IC-BNMT 2010, and IET CCWMC 2009. He was
awarded the First Grade Award of Technological Invention Award in
Ministry of Education of China, and the Second Grade Award of
Scientific \& Technical Progress from China Institute of
Communications.
\end{IEEEbiographynophoto}

\end{document}